\documentclass[a4paper,showpacs]{revtex4}
\usepackage{graphicx}
\usepackage{amsmath}
\usepackage{amssymb}
\usepackage{amsfonts}

\begin{document}

\title{Reconstructing cosmic acceleration from f(R) modified gravity}

\author{Antonio Jes\'us L\'opez-Revelles}

\affiliation{Consejo Superior de Investigaciones Cient\'{\i}ficas,
ICE(CSIC-IEEC), Campus UAB Facultat de Ci\`encies, Torre C5-Parell-2a
pl, E-08193 Bellaterra (Barcelona) Spain}


\begin{abstract}
 A variant of the accelerating cosmology reconstruction program is developed for $f(R)$ gravity. Reconstruction schemes in terms of e-foldings 
and by using an auxiliary scalar field are developed. An example of a model with a transient phantom behavior without real matter is explicitly 
discussed in both schemes. Detailed comparison of the two schemes of reconstruction is presented. It seems to support physical non-equivalence 
of the two frames.
\end{abstract}

\maketitle


\section{INTRODUCTION}

	One of the most important problems of modern cosmology is the explanation of the current universe speed-up, first discovered in 
\cite{4,5}. A convenient way to express this situation is to introduce a new form of energy, called Dark Energy (DE). During the last few 
years several theories have been developed in order to formulate and try to explain the dark energy universe.

	As gravitational alternatives for DE, modified gravity theories have been formulated, calling for plausible late-time modification of 
General Relativity (GR). Many modified gravity models have been proposed in the literature (for a review, see \cite{1}), starting from the very 
simple $1/R$ theory \cite{16,10,11} to more elaborated ones.

	For any such theory to be valid it is always strictly required that it accurately describes the known sequence of cosmological epochs, 
specifically it must accurately fit an increasing number of more and more precise observational data \cite{14,15}. Another important issue to 
be taken into account concerns the different types of finite-time, future cosmological singularities (see \cite{29}). It has been demonstrated 
(see \cite{32, 33}) that in modified gravities all four known types of finite-time future singularities may possibly appear. However, modified 
gravity can have the chance to cure all of these future singularities (see \cite{31,aj2}).

	Several reconstruction methods that make possible to reproduce any given cosmology have appeared for $f(R)$ modified gravity 
\cite{2,aj1,3}. In the following these methods will be shown. 

\section{COSMOLOGICAL RECONSTRUCTION OF REALISTIC MODIFIED $f(R)$ GRAVITIES IN TERMS OF E-FOLDINGS}

	In this section, the techniques of \cite{2,aj1} are summarized in order to construct an $f(R)$ model realizing any given cosmology. The 
	starting action for $f(R)$ gravity (see e.g.~\cite{1}, for a general review) is
		\begin{equation}\label{t1}
            		S = \int d^4 x \sqrt{-g} \left( \frac{f(R)}{2 \kappa^2} + \mathcal{L}_{matter} \right).
		\end{equation}

	Using the e-folding variable, $N = \ln{\frac{a}{a_0}}$, assuming the matter density $\rho$ is given in 
	terms of a sum of fluid densities with constant EoS parameters, $\omega_i$, and considering the Hubble rate is given by 
	$G(N) \equiv  H^2$, the first Friedmann equation for the action (\ref{t1}) yields 
		\begin{equation}\label{t6}
			{\textstyle  0 = - 9 G(N(R)) \left[ 4 G'(N(R)) + G''(N(R)) \right] \frac{d^2 f(R)}{dR^2} + \left[ 3 G(N(R)) + \frac{3}{2} G'(N(R)) 
			\right] \frac{d f(R)}{dR} - \frac{f(R)}{2} +}$$
			$${\textstyle  + \sum \limits_i \rho_{i0} a_0^{-3 (1 + \omega_i)} e^{-3 (1 + \omega_i) N(R)}}.
		\end{equation}
	This is a differential equation for $f(R)$, where the variable is the scalar curvature $R$, which is here $R = 3 G'(N) + 12 G(N)$
	(for detailed calculations see \cite{2}).\\

		Let us consider an evolution given by the following Hubble parameter:
			\begin{equation}\label{t7}
				{\textstyle  H^2(N) = H_0 \ln{\left( \frac{a}{a_0} \right)} + H_1 = H_0 N + H_1 = G(N)},
			\end{equation}
		where $H_0$ and $H_1$ are positive constants. We can, in this case, achieve a transient phantom behavior without the presence 
		of real matter. Indeed, from $R = 3 G'(N)+ 12 G(N)$, we can obtain $N = N(R)$ and Eq.~(\ref{t6}) takes the form (for more 
		details, see \cite{aj1})
			\begin{equation}\label{t9}
				{\textstyle  0 = - 3 H_0 (R - 3 H_0) \frac{d^2 f(R)}{dR^2} + \left( \frac{R - 3 H_0}{4} + \frac{3 H_0}{2} \right) 
				\frac{d f(R)}{dR} - \frac{f(R)}{2}},
			\end{equation}
		whose solutions are given by the Kummer's series $\Phi (a, b; z)$, namely
			\begin{equation}\label{t11}
				{\textstyle  f(R) = C \, \Phi \left( -2, - \frac{1}{2}; \frac{R - 3 H_0}{12 H_0} \right) = C \left( - \frac{1}{4} + 
				\frac{1}{2 H_0} R - \frac{1}{36 H_0^2} R^2 \right)},
			\end{equation}
		where $C$ is a constant. As a consequence, with this $f(R)$ theory, as given by Eq.~(\ref{t11}), we can reproduce the transient 
		phantom behavior without real matter given by Eq.~(\ref{t7}).

\section{COSMOLOGICAL RECONSTRUCTION OF REALISTIC MODIFIED $f(R)$ GRAVITIES USING A SCALAR FIELD}
	
	In this section it will be used the technique of \cite{aj1,3} in order to construct an $f(R)$ model realizing any given cosmology. We 
	start from the action for modified gravity
		\begin{equation}\label{t12}
			S = \int d^4 x \sqrt{-g} (f(R) + \mathcal{L}_{matter}),
		\end{equation}
	which is equivalent to 
		\begin{equation}\label{t13}
			S = \int d^4 x \sqrt{-g} (P(\phi) R + Q(\phi) + \mathcal{L}_{matter}).
		\end{equation}
	Here, $\mathcal{L}_{matter}$ is the matter Lagrangian density and $P$ and $Q$ are proper functions of the auxiliary scalar field, 
	$\phi$. After some calculations and some assumptions (see \cite{3}) we can obtain the following differential equation for the function 
	$P(\phi)$
		\begin{equation}\label{t23}
			0 = 2 \frac{d^2 P(\phi)}{d\phi^2} - 2 g'(\phi) \frac{d P(\phi)}{d\phi} + 4 g''(\phi) P(\phi) + \sum \limits_i (1 + 
			\omega_i) \rho_{i0} a_0^{-3 (1 + \omega_i)} e^{-3 (1 + \omega_i) g(\phi)}.
		\end{equation}
	While the function $Q(\phi)$ is given by
		\begin{equation}\label{t24}
			Q(\phi) = - 6 (g'(\phi))^2 P(\phi) - 6 g'(\phi) \frac{d P(\phi)}{d\phi} + \sum \limits_i \rho_{i0} a_0^{-3 (1 + 
			\omega_i)} e^{-3 (1 + \omega_i) g(\phi)}.
		\end{equation}
	The function $f(R)$ can be calculated as $f(R) = P(\phi(R)) R + Q(\phi(R))$. Thus, any given cosmology, expressed as 
	$a(t) = a_0 e^{g(t)}$, can be realized by some specific $f(R)$-gravity.\\

		We consider another time the transient phantom behavior, without real matter, given by (\ref{t7}). In this case, Eq.(\ref{t23}) 
		reduces to
			\begin{equation}\label{t27}
				0 = \frac{d^2 P(\phi)}{d\phi^2} - \frac{H_0}{2} (\phi - c) \frac{d P(\phi)}{d\phi} + H_0 P(\phi).
			\end{equation}
		We can obtain $P(\phi)$ from this equation and $Q(\phi)$ from (\ref{t24}).The function $f(R) = P(\phi(R)) R + Q(\phi(R))$
		reduces to (for more details, see \cite{aj1}):
			\begin{equation}\label{t32}
				{\textstyle f(R) = C_1 \left[ A + H_0 \sqrt{3(R - 3 \sqrt{2 H_0})} - \frac{R^2}{24} \right] + C_2 \left\{ 
				\left[ A - \frac{R^2}{24} \right] \left( \frac{\sqrt{\frac{3(R - 3 \sqrt{2 H_0})}{2 H_0}} \, e^{\frac{R - 3 
				\sqrt{2 H_0}}{24}}}{24 - 2(R - 3 \sqrt{2 H_0})} - \right. \right. }$$
				$${\textstyle \left. \left. - \frac{i}{4 \sqrt{H_0}} \int \limits_0^{\frac{i}{2} \sqrt{\frac{R}{6} - 
				\sqrt{\frac{H_0}{2}}}} e^{-y^2} dy \right) - \frac{3}{4} \sqrt{2 H_0} e^{\frac{R - 3 \sqrt{2 H_0}}{24}} - 
				\frac{i \sqrt{3 H_0}}{4} \sqrt{R - 3 \sqrt{2 H_0}} \int \limits_0^{\frac{i}{2} \sqrt{\frac{R}{6} - 
				\sqrt{\frac{H_0}{2}}}} e^{-y^2} dy \right\}},
			\end{equation}
		where ${\textstyle A = 3 \left( \sqrt{\frac{H_0}{2}} + \frac{H_0}{4} \right)}$. We thus prove that, with this scheme, we are 
		able to obtain the $f(R)$ model (\ref{t32}) which reproduces the desired transient phantom behavior without real matter, as 
		given by (\ref{t7}).

\section{SUMMARY AND DISCUSSION}

	In summary, two different schemes of reconstruction have been developed for modified gravity. With these methods, any explicitly given 
	cosmology can be realized as a corresponding modified gravity.

	The same example has been developed in the two cases. The result obtained in the first scheme (in terms of e-foldings) is given by 
	Eq.(\ref{t11}) while for the second scheme (using an auxiliary scalar field), the $f(R)$ obtained is given by (\ref{t32}). As one can 
	easily see, the results obtained for both methods are in fact different. The reason behind this is the fact that action (\ref{t13}) 
	corresponds to a wider class of theories than the action (\ref{t12}) (for a related and quite detailed discussion, see \cite{27,28}). 
	Nevertheless, if in Eq.~(\ref{t32}) we set $C_2 = 0$, then the results coming from both schemes are similar, at least in the sense that, 
	for low curvatures, they behave as constant, while for large curvatures the behavior is in both cases as $R^2$.

	Finally, I want to remark that this work is based on the work done with Emilio Elizalde, which was published in Physical Review D (see 
	\cite{aj1}).

\section*{Acknowledgements}
  I thank Emilio Elizalde and Sergei Odintsov for suggesting the problem and giving the ideas to carry out this task. The work has been 
supported by MICINN (Spain), project FIS2006-02842, and by AGAUR (Generalitat de Ca\-ta\-lu\-nya), contract 2009SGR-994. I also acknowledge 
a JAE fellowship from CSIC.


\end{document}